\documentclass[aps,prl,twocolumn,superscriptaddress,amsfont,graphicx,nofootinbib,preprintnumbers]{revtex4-1}%
\usepackage{color,graphicx,epsfig}
\usepackage{ifpdf}
\usepackage{amsmath}
\usepackage{bm}
\usepackage{color}
\usepackage[english]{babel}
\usepackage{graphicx}%
\usepackage{amsfonts}%
\usepackage{amssymb}
\usepackage{braket}
\usepackage{hyperref}
\usepackage{enumerate}
\usepackage{comment}
\usepackage[dvipsnames]{xcolor}

\definecolor{nicered}{rgb}{0.7,0.1,0.1}
\definecolor{nicegreen}{rgb}{0.1,0.5,0.1}
\hypersetup{colorlinks,citecolor= nicegreen,linkcolor= nicered}

\usepackage{graphicx}
\usepackage{dcolumn}
\usepackage{bm}
\usepackage{slashed}

\begin{document}

\title{Atmospheric Dark Matter and Xenon1T Excess}

\author{Liangliang Su}
\affiliation{Department of Physics and Institute of Theoretical Physics, Nanjing Normal University, Nanjing, 210023, China}
\affiliation{School of Physics, Yantai University, Yantai 264005, China}

\author{Wenyu Wang}
\affiliation{Faculty of Science, Beijing University of Technology, Beijing, China}

\author{Lei Wu}
\affiliation{Department of Physics and Institute of Theoretical Physics, Nanjing Normal University, Nanjing, 210023, China}

\author{Jin Min Yang}
\affiliation{CAS Key Laboratory of Theoretical Physics, Institute of Theoretical Physics, Chinese Academy of Sciences, Beijing 100190, China}
\affiliation{School of Physics, University of Chinese Academy of Sciences, Beijing 100049, China}

\author{Bin Zhu}
\affiliation{School of Physics, Yantai University, Yantai 264005, China}
\affiliation{Department of Physics, Chung-Ang University, Seoul 06974, Korea}

\date{\today}

\begin{abstract}
Very recently, the Xenon1T collaboration has reported an intriguing electron recoil excess, which may imply for light dark matter. In order to interpret this anomaly, we propose the atmospheric dark matter (ADM) from the inelastic collision of cosmic rays (CRs) with the atmosphere. Due to the boost effect of high energy CRs, we show that the light ADM can be fast-moving and successfully fit the observed electron recoil spectrum through the ADM-electron scattering process. Meanwhile, our ADM predicts the scattering cross section $\sigma_e \sim {\cal O}(10^{-38}- 10^{-39}$) cm$^{2}$, and thus can evade other direct detection constraints. The search for light meson rare decays, such as $\eta \to \pi + \slashed E_T$, would provide a complementary probe of our ADM in the future.
\end{abstract}
\pacs{Valid PACS appear here}
\maketitle


\section{Introduction}

The existence of dark matter (DM) has been established in cosmological and astrophysical experiments. Besides the gravitational effects, other possible interactions of dark matter are still unknown. The various searches for dark matter in the direct detections, indirect detections and collider experiments are in progress, albeit no convincing signals have been observed. In particular, the direct detections~\cite{Jungman:1995df} that aim for Weakly Interacting Massive Particle (WIMP)~\cite{Lee:1977ua} have reached great sensitivities, which are approaching to the irreducible neutrino floor. Their null results produce very stringent limits on the WIMP DM-nucleus scattering cross section, and lead to a shift of focus towards light dark matter particles~\cite{Bertone:2018krk,Knapen:2017xzo}.

As the average velocity of dark matter is around $10^{-3}c$ in the Milky Way halo, the sensitivity of traditional direct detections that measure nuclear recoils rapidly decreases for DM mass below $\sim 1$ GeV. In order to access the sub-GeV dark matter, many new techniques and new types of detectors have been proposed (see e.g.~\cite{Hochberg:2015pha,Schutz:2016tid,Giudice:2017zke,Ibe:2017yqa,Dolan:2017xbu,Smirnov:2020zwf,Zhang:2020nis}). Among them, search for DM scattering off electrons has been demonstrated to be a useful way of improving the discovery potential of light dark matter. Concretely, since the electron is bounded to the atom, it can have a non-negligible momentum. When the DM particles scatter off these high-momentum electrons, the xenon atoms can be ionized in the liquid target. In this process, the energy transfer to the detector is about $E_R \sim$ keV. Then such
ionization (and scintillation) signals can be detectable at the dual-phase liquid Xenon detectors.

Very recently, with an exposure of 0.65 tonne-years and an unprecedentedly low background, the Xenon1T collaboration has reported an about $3.5\sigma$ excess of events in the electron recoil range of 1 keV $<E_R<$ 7 keV with 285 events over the backgrounds of $232\pm15$ events~\cite{Aprile:2020tmw}. The main excess events appear in the 2-3 keV bins, while other bins are approximately consistent with the expected background events. In the analysis of Xenon1T, it is pointed out that the observed electron recoil spectrum can be fitted by the solar axion~\cite{vanBibber:1988ge,Moriyama:1995bz,Redondo:2013wwa} with an axion-electron coupling $g_{ae} \simeq 3.7 \times 10^{-12}$, which, however, is in tension with the stellar cooling constraint, $g_{ae} \lesssim 0.3 \times 10^{-12}$~\cite{Giannotti:2017hny}. Other speculations about this excess have been discussed in~\cite{Aprile:2020tmw,Takahashi:2020bpq,Kannike:2020agf,Fornal:2020npv,Boehm:2020ltd,Alonso-Alvarez:2020cdv,Chen:2020gcl,Du:2020ybt}. Although the possibility of contamination from $\beta$ decay of tritium is not excluded, such an anomaly is still intriguing and may be a sign of light dark matter.

\begin{figure}[ht]
  \centering
  \includegraphics[height=7cm,width=7cm]{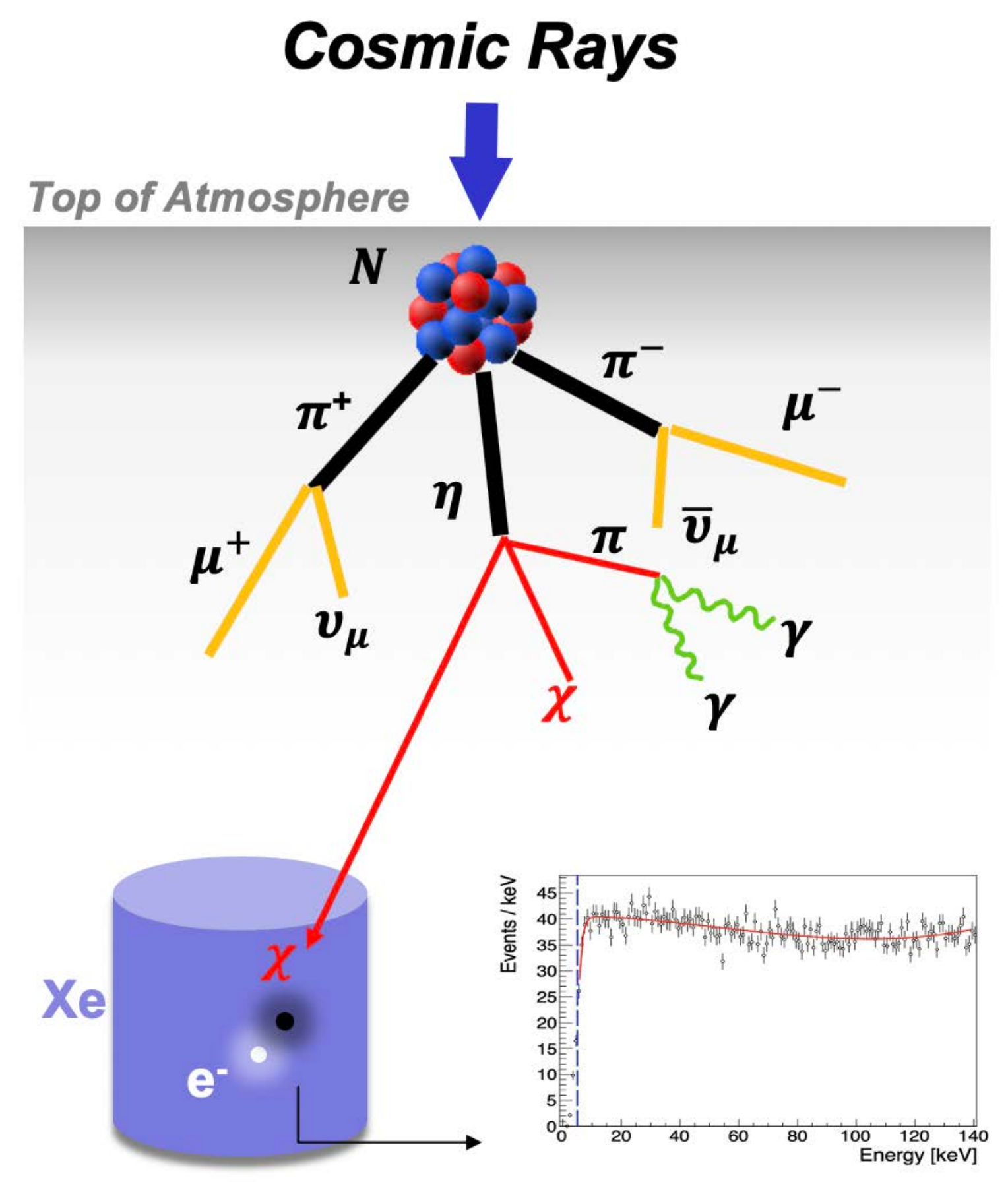}
  \caption{Atmospheric dark matter from inelastic collision of CRs and atmosphere for Xenon1T electron recoil excess. Here $\chi$ stands for the dark matter particles.}\label{attenuation}
\end{figure}

In general, the light dark matter bounded in galaxies moves with a low velocity and it is impossible to fit this Xenon1T electron data because of the small recoil energy. However, there are several astrophysical processes that can accelerate the dark matter to have velocities much higher than its galactic escape velocity~\cite{Kouvaris:2015nsa,An:2017ojc,Emken:2017hnp,Cappiello:2018hsu}. This kind of fast-moving dark matter will scatter with nucleus or electron of target in the direct detection, and produce the detectable signals. For example, the light dark matter can be boosted to (semi-)relativistic speeds through its elastic scattering with the high energy cosmic rays (CRs)~\cite{Bringmann:2018cvk,Bondarenko:2019vrb,Dent:2019krz,Wang:2019jtk,Plestid:2020kdm,Ge:2020yuf}. In this paper, we propose the light boosted dark matter produced in the inelastic collision of CRs with the atmosphere (c.f. Fig.1) to explain the Xenon1T excess. Different from the up-scattering mechanism, this scheme is independent of the density of pre-existing dark matter, and thus naturally provides a sufficient source of boosted dark matter. Besides, the Xenon1T electron data requires the DM to interact with the electron as well. In order to reconcile the tension between DM-nucleus and DM-electron scattering cross sections, we introduce a scalar and a vector mediator, which couples with quark and electron, respectively.

\section{Atmospheric Dark Matter}
The main components of high energy CRs are protons and heliums, which can inelastically collide with the interstellar medium or the atmosphere on Earth. The latter is usually the dominant source of the energetic dark matter~\cite{Alvey:2019zaa}. For simplicity, we assume the protons as the incoming cosmic ray flux and parameterize it as in Ref.~\cite{Boschini:2017fxq}. The differential cosmic ray flux $d\phi_p(T_p,h)/dT_p$ is the function of proton energy $T_p$ and the height $h$ from the ground level, which will be diluted as traveling through the atmosphere,
\begin{equation}
\frac{\mathrm{d}}{\mathrm{d} h} \frac{\mathrm{d} \phi_{p}\left(T_{p}, h\right)}{\mathrm{d} T_{p}}=\sigma_{p N}(T_p) n_{N}(h) \frac{\mathrm{d} \phi_{p}\left(T_{p}, h\right)}{\mathrm{d} T_{p}}.
\label{eqn:flux}
\end{equation}
Here we assume the nitrogen as nuclei target in the atmosphere. $\sigma_{pN}$ is the inelastic proton-nitrogen cross section and $n_N$ is the number density of nitrogen. The initial value of the flux is evaluated at $h_{\max}=180 {\rm km}$. It should be noted that the primary cosmic ray will generate secondary particles with lower energy, such as nucleons, pions and kaons, via the interactions in the atmosphere or in the earth. In principle, the secondary cosmic ray collisions may also become the sources of our fast-moving DM (assuming the boosted effect can be still large enough, e.g. $v_\chi \sim 0.1c$~\cite{Kannike:2020agf}). The numerical estimation of these contributions would need to solve the transport equations~\cite{Lipari:1993hd,Gamez:2019dex} by considering the additional invisible decay of $\eta$ meson. We follow  \cite{Alvey:2019zaa} to neglect these re-generations and secondary scatterings involved in a detailed cosmic ray shower model, and other higher order effects. This will lead to a smaller flux and allow us to obtain a conservative result. Including these secondary contributions would make our fit easier.

Since the inelastic proton-nitrogen cross section is approximately constant in the relevant energy range, we can absorb the $h$-dependence of $\phi_p$ into a dilution factor $y_p(h)$ for simplicity,
\begin{equation}
\frac{\mathrm{d} \phi_{p}\left(T_{p}, h\right)}{\mathrm{d} T_{p}}=y_p(h) \frac{\mathrm{d} \phi_{p}\left(T_{p}, h_{\max }\right)}{\mathrm{d} T_{p}},
\label{eqn:definition}
\end{equation}
where we set the boundary condition of suppression factor as $y_p|_{h_{\max}=180\mathrm{km}}=1$. Then, we can substitute the Eq.~\ref{eqn:definition} into suppression function~Eq.\ref{eqn:flux} and yields,
\begin{equation}
\frac{\mathrm{d} y_p(h)}{\mathrm{d} h}=\sigma_{p N} n_{N}(h) y_p(h).
\end{equation}
After integration over the height, we can obtain the dilution factor,
\begin{equation}
y_p(h)=\exp \left(-\sigma_{p N} \int_{h}^{h \max } \mathrm{d} \tilde{h} n_{N}(\tilde{h})\right).
\end{equation}
In the numerical calculation, we simulate the collision of incoming CRs with the nitrogen via the process $pN \to X$ by the package {\textsf{CRMC}}~\cite{Pierog:2013ria,Baus:2019,alvey:2019}, where $X$ denotes the meson produced in this inelastic collision. Then, these mesons will decay to the on-shell dark matter mediator $M$ plus the SM particles, such as $\eta \to \pi M$, which is followed by the two-body decay $M \to \chi \bar{\chi}$. 

Such a scenario can be realized in the hadrophilic dark sector~\cite{Batell:2018fqo},
\begin{align}
\mathcal{L}_S&=i \bar{\chi}\left(\gamma^{\mu}\partial_{\mu}-m_{\chi}\right) \chi+\frac{1}{2} \partial_{\mu} S \partial^{\mu} S\nonumber\\
&-\frac{1}{2} m_{S}^{2} S^{2}-\left(g^S_{\chi} S \bar{\chi}_{L} \chi_{R}+g_{u} S \bar{u}_{L} u_{R}+\mathrm{h.c.}\right),
\end{align}
where a scalar mediator $S$ only couples to up-quark. The masses of hadrophilic mediator and dark matter are donated by $m_S$ and $m_\chi$, while the couplings of mediator $S$ with the dark matter and up-quark are denoted by $g^S_\chi$ and $g_u$, respectively. Then, the resulting branching ratio of the new $\eta$ decay is given by
\begin{equation}
\operatorname{Br}\left(\eta \rightarrow \pi^{0} S\right)=\frac{c_{S \pi^{0} \eta}^{2} g_{u}^{2} B^{2}}{16 \pi m_{\eta} \Gamma_{\eta}} \lambda^{1 / 2}\left(1, \frac{m_{S}^{2}}{m_{\eta}^{2}}, \frac{m_{\pi^{0}}^{2}}{m_{\eta}^{2}}\right).
\label{etadecay}
\end{equation}
where the kinematic function $\lambda(a,b,c)=a^2+b^2+c^2-2ab-2ac-2bc$ and $B=m^2_\pi/(m_u+m_d)\simeq 2.6$ GeV. The coefficient $C_{S\pi^0\eta}=(\frac{1}{\sqrt{3}}\cos\theta-\sqrt{\frac{2}{3}}\sin\theta)|_{\theta=-20^0}$ is the mixing parameter of $\eta$ and $\eta^\prime$~\cite{Gan:2020aco}.

\begin{figure}[ht]
  \centering
  \includegraphics[height=8cm,width=8.5cm]{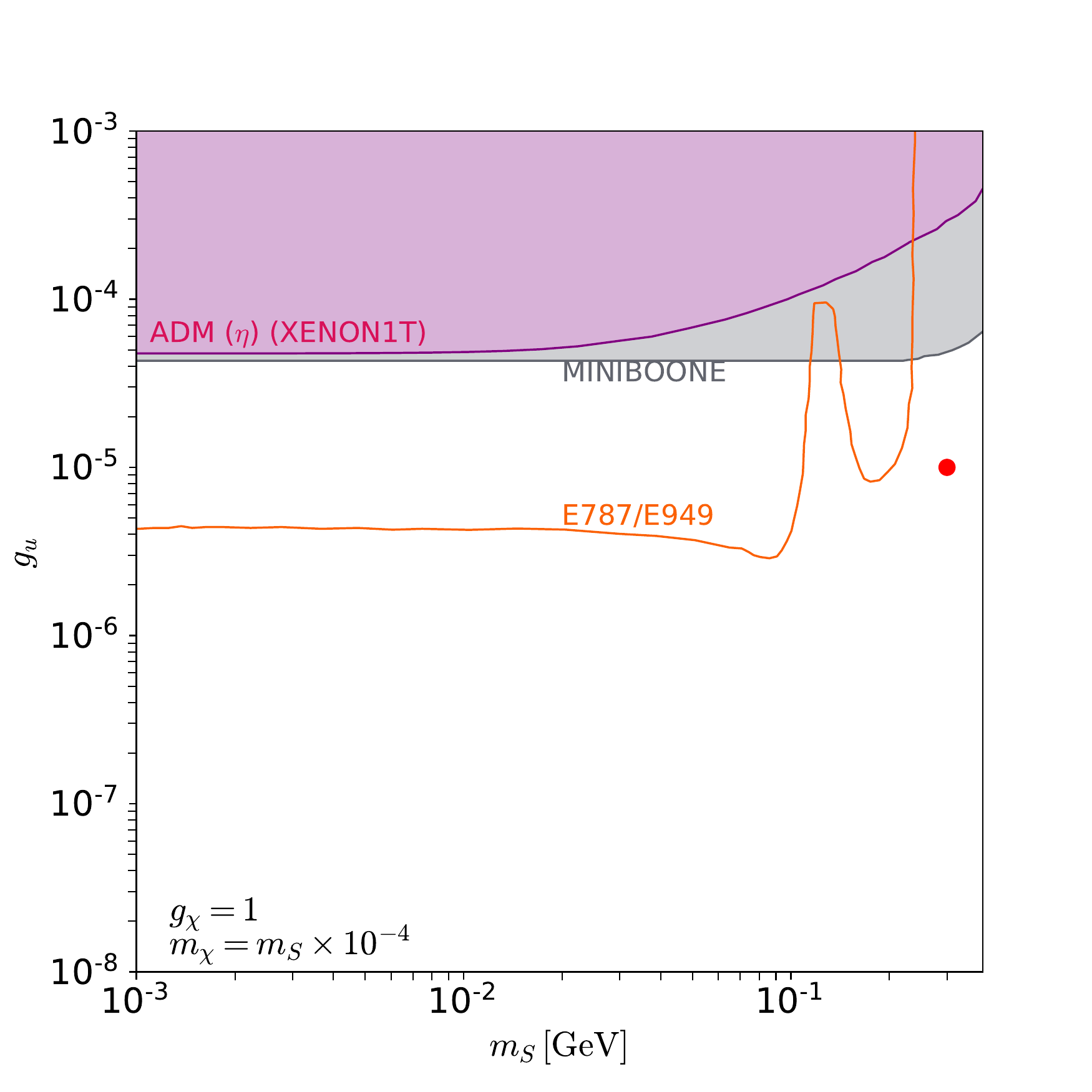}
  \caption{The exclusion limits on the plane of the coupling $g_u$ versus the scalar mediator mass $m_S$, where the coupling $g_\chi=1$ and $m_\chi=m_S\times 10^{-4}$. The benchmark point (red bullet) is taken as $m_\chi=30$ keV, $m_S =300$ MeV, $g_u=10^{-5}$ and $g_\chi=1$.}\label{hadrophilic}
\end{figure}
Since our mediator $S$ is required to be produced on-shell, we focus on the mass range of $m_S < m_\eta-m_\pi$. There is a strong constraint on $g_u$ and $m_S$ from the decay of kaon meson~\footnote{It should be noted that the neutral pion decay is usually a sensitive way of searching for light DM, such as $\pi \to \nu\bar{\nu}\gamma$. However, due to the parity conservation, this bound is not applicable to our scalar mediator.}, such as $K \to \pi S (\to \chi\chi)$. Using the result of search for $K\to \pi \nu\bar{\nu}$ in E787/949 experiment~\cite{Adler:2002hy,Adler:2004hp,Adler:2008zza,Artamonov:2009sz}, we derive the limit on the plane of $g_u$ and $m_S$ in Fig.~\ref{hadrophilic}, where we assume $g_\chi=1$ and $m_\chi=m_S \times 10^{-4}$. It can be seen that $g_u$ should be less than about $4\times 10^{-6}$ as $m_S<m_\pi$. While in the mass range of $m_\pi<m_S< m_K-m_\pi$, $g_u$ is allowed to be larger. It should be noted that the conventional beam dump experiments of searching for the decay products of new particles that are produced in fixed-target collisions and decay far downstream are not applicable because our mediator $S$ dominantly decays to DM. However, our ADM can be produced by the $\eta$ meson decay in primary collisions of protons in a beam dump, and then elastically scatters with nucleon. Hence, the null result of MINIBooNE experiment~\cite{Aguilar-Arevalo:2018wea} gives a upper bound $g_u \lesssim 4 \times 10^{-5}$, which is stronger than the limit derived from the Xenon1T data of the spin-independent DM-nucleus scattering. On the other hand, due to the huge QCD backgrounds, there is no limit on the process $pp \to \eta \to \pi^0(\to \gamma\gamma) + \slashed E_T$ in the LHC experiment. But it is found that the search for the mono-jet events from the process $pp \to gS(\to \chi\chi)$ will require $g_u <0.1$~\cite{Batell:2018fqo}. Therefore, in order to avoid the above constraints and produce the sufficient flux to explain the Xenon1T electron recoil excess, we take $g_u=10^{-5}$ and $m_S=300$ MeV as our benchmark point (the red bullet in Fig.~\ref{hadrophilic}), which corresponds to $Br(\eta \to \pi \chi \bar{\chi}) \simeq 1\times10^{-5}$.

After produced from the decays of mesons, the flux of ADM will be further attenuated by the secondary scattering in the earth. Similar to the above, the attenuation factor of dark matter can be written as
\begin{equation}
y_{d}(h, \theta, \phi)=\exp \left(-\sigma_{\chi N} \int_{0}^{l_d} \mathrm{d}z~n(r(z)-R_E) \right)
\end{equation}
where $n$ is the number density of nucleus, and $R_E=6378.1$ km is the value of Earth radius and $l_d$ denotes the line of sight distance between the point of dark matter production and the detector,
\begin{eqnarray}
\ell_{d}^{2}(h, \theta)&=&\left(R_{E}+h\right)^{2}+\left(R_{E}-h_{d}\right)^{2} \nonumber \\ && -2\left(R_{E}+h\right)\left(R_{E}-h_{d}\right) \cos \theta
\end{eqnarray}
where $h_d=1.4$ km is the depth of the detector, and $\theta$ is the angle between the point of dark matter production and the detector. $\sigma_{\chi N}$ is the elastic cross section between dark matter and nucleus. Due to attenuation effects, the flux of ADM will exponentially decrease as the DM-nucleon spin-independent scattering cross section $\sigma^{SI}_\chi \gtrsim 10^{-28}$ cm$^{2}$. In our numerical calculations, we assume $\sigma^{SI}_\chi=10^{-34}$ cm$^{2}$, which can escape the constraint from the null result of Xenon1T spin-independent DM-nucleon scattering~\cite{Alvey:2019zaa}.

Combining the dilution factor of cosmic ray $y_p$ and the attenuation factor of dark matter $y_d$, we can obtain the differential flux of ADM at the depth of $h_d$ below the surface of the Earth,
\begin{eqnarray}
\frac{\mathrm{d} \phi_{\chi}^{h_d}}{\mathrm{d} T_{\chi}}=G \int_{T_{p}^{\mathrm{min}}}^{T_{p}^{\max }} \mathrm{d} T_{p} \frac{1}{\Omega\left(T_{p}\right)} \frac{\mathrm{d} \phi_{p}\left(h_{\mathrm{max}}\right)}{\mathrm{d} T_{p}} \frac{\mathrm{d} \sigma_{p N \rightarrow \chi \chi \gamma}}{\mathrm{d} T_{\chi}}
\end{eqnarray}
with the geometrical factor
\begin{eqnarray}
G&=&\int_{0}^{h_{\mathrm{max}}} \mathrm{d} h\left(R_{E}+h\right)^{2} \int_{0}^{2 \pi} \mathrm{d} \phi \int_{-1}^{+1} \mathrm{d} \cos \theta \nonumber \\ &&\cdot \frac{y_{d}\left(h, \theta ; \sigma_{\chi N}\right) y_{p}(h, \theta)}{\ell_{d}^{2}(h, \theta)} n_{N}(h),
\end{eqnarray}

The inelastic differential cross section is given by
\begin{eqnarray}
\frac{d\sigma_{pN \to \pi \chi \bar{\chi}}}{dT_{\chi}}&=&\frac{d\sigma_{pN}}{dT_{\chi}} \frac{\Gamma_{\pi \chi \bar{\chi}}}{\Gamma_{\mathrm{tot}}}\nonumber \\&\simeq& \frac{\sigma_{pN}}{T_{\chi}^{\max}} \mathrm{BR}(\eta \to \pi \chi\bar{\chi}).
\end{eqnarray}
Here we assume an isotropic scattering and take a uniform distribution of the ADM kinetic energy.

As mentioned above, only the hadrophilic scalar mediator cannot account for Xenon1T electron detection. Therefore we introduce additional leptophilic vector mediator $A'$ to communicate the dark matter and electrons,
\begin{eqnarray}
\mathcal{L}_{A'}=g_e\bar{e}\gamma^{\mu}e A_{\mu}^{\prime}+g^{A'}_{\chi}\bar{\chi}\gamma^{\mu}\chi A_{\mu}^{\prime}+m^2_{A'}A^{\prime\mu} A^{\prime}_{\mu},
\end{eqnarray}
where $g^{A'}_\chi$ and $g_e$ are the couplings of mediator $A^{\prime}$ with the dark matter and electron, respectively. $m_{A'}$ is the mass of mediator. Then we can calculate the differential recoil rate by
\begin{equation}
\frac{dR}{dE_R}=\epsilon(E_R)  n_T\int_{T_{\chi}^{\min}}^{T_{\chi}^{\max}}\frac{d\phi_{\chi}}{dT_{\chi}}\frac{d\sigma_{\chi e}}{dE_R},
\label{rate}
\end{equation}
where $n_T=4.2\times 10^{27}$ is number density of Xenon per tonne and $\epsilon(E_R)$ is detection efficiency~\cite{Aprile:2020tmw}. The ADM flux $d\phi^{h_d}/dT_{\chi}$ is given by Eq.~\ref{flux}. For a fixed DM velocity, the differential cross section of the ADM scattering with the electron can be written as,
\begin{equation}
\frac{\mathrm{d} \sigma_{\chi e}}{\mathrm{d} E_{R}}=\frac{\sigma_{e}}{2 m_{\mathrm{e}} v_{\chi}^{2}} \sum_{nl}\int_{q_{\min}}^{q_{\max}} a_{0}^{2} q \mathrm{d} q|F(q)|^{2} K_e(E_R,q).
\label{cxer}
\end{equation}
where $a_0=1/(\alpha m_e)$ is the Bohr radius. $\sigma_e$ is the DM-free electron scattering cross section at the given momentum transfer $q=1/a_0$. We attribute the momentum-dependent effect into the dark matter form factor $F(q)$. In non-relativistic limit, the form factor is $F(q)=(\alpha^2 m^2_e + m^2_{A^{\prime}})/(q^2+m^2_{A^{\prime}})$. While for boosted dark matter, it will rely on the mediator mass $m_M$ and the kinetic energy of dark matter $T_{\chi}$,
\begin{eqnarray}
\left|F(q)\right|^{2}&=&\frac{\left(\alpha^{2} m_{e}^{2}+m_{A^{\prime}}^{2}\right)^{2}}{\left(q^{2}+m_{A^{\prime}}^{2}\right)^{2}}  \times \frac{1}{2 m_{e} m_{\chi}^{2}} \nonumber \\ && \times [ 2 m_{e}\left(m_{\chi}+T_{\chi}\right)^{2}-E_{R}(m_{\chi}+m_{e})^{2} \nonumber \\ && +2 m_{e} E_{R}T_{\chi} +m_{e} E_{R}^{2} ]
\end{eqnarray}
It can be seen that such a factor will behave like $(T_{\chi}/m_{\chi})^2$, and thus can be much larger than 1 when $T_{\chi} \gg m_{\chi}$. $K_e$ is atomic form factor which is summation of all possible energy levels of ionization factor,
\begin{equation}
    K_e(E_R,q)=\frac{\alpha^2 m_e}{4E_R}\frac{(m_e+m_{\chi})^2}{m_{\chi}^2}\sum_{n,l}\left|f_{nl}(\sqrt{2m_e E_R},q)\right|^2
\end{equation}
 By using the given bound wave functions and unbound wave functions, this form factor for electron in the different shells is calculated in \cite{Essig:2017kqs,Roberts:2016xfw}. The limits of integration of the momentum transfer $q_{\min}$ and $q_{\max}$ are given by,
\begin{equation}
q_{\min,\max}=m_{\chi} v_{\chi} \mp \sqrt{m_{\chi}^{2} v_{\chi}^{2}-2 m_{\chi} E_{R}},
\label{qtransfer}
\end{equation}
where $v_\chi$ is the velocity of ADM in its scattering with electron. 

\section{Xenon1T Electron recoil Excess}

\begin{figure}[ht]
  \centering
  \includegraphics[height=8cm,width=8.5cm]{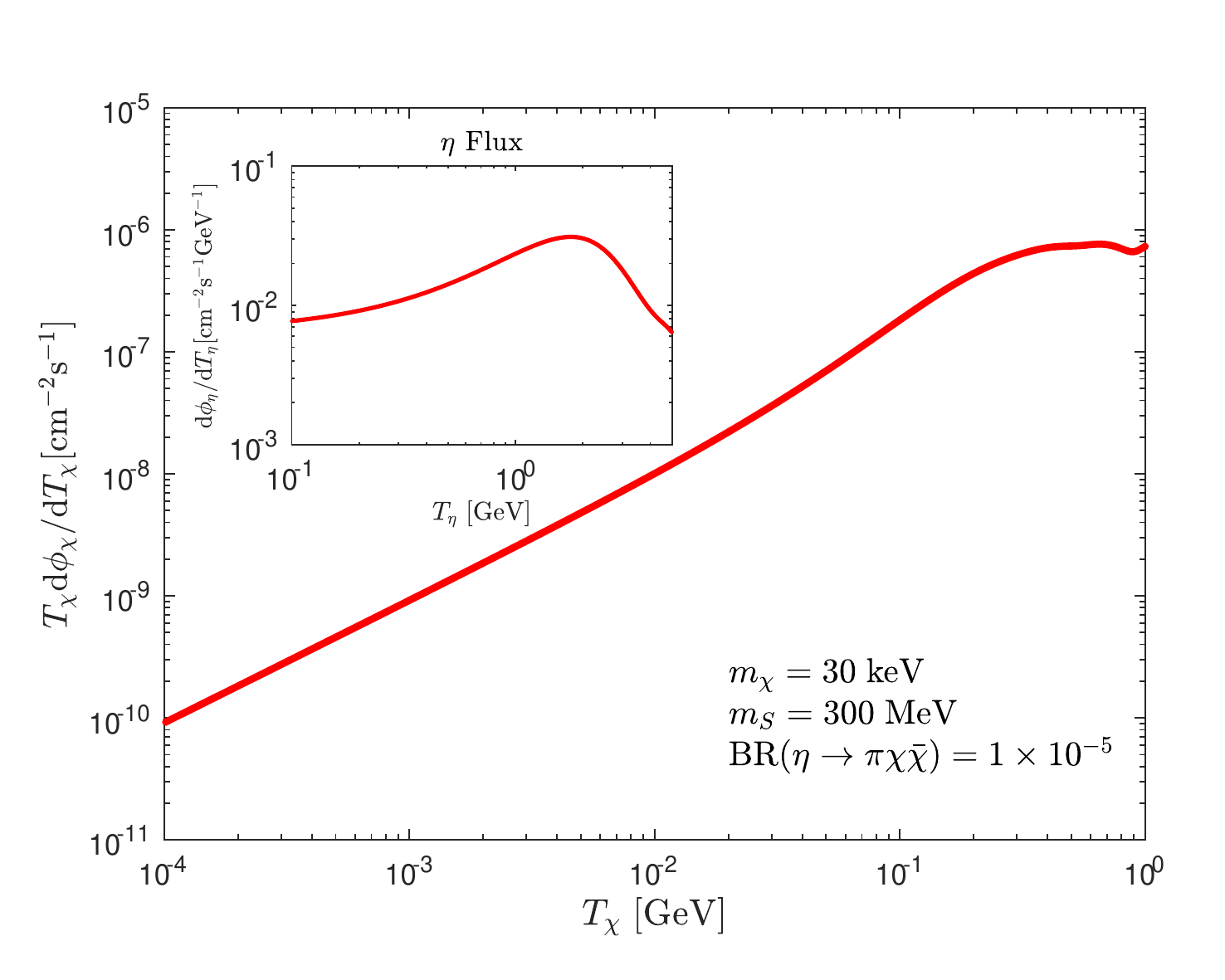}
  \caption{The expected flux of ADM and $\eta$ meson. The benchmark point is the same as that in Fig.~\ref{hadrophilic}. The nucleon spin-independent cross section $\sigma^{SI}_\chi=10^{-34}$ cm$^{2}$ and the branching ratio $Br(\eta \to \pi \chi\bar{\chi})=1\times10^{-5}$ are assumed.}\label{flux}
\end{figure}
In Fig.~\ref{flux}, we show the flux of ADM  and $\eta$ meson. It can be seen that the flux of ADM has a peak in the (semi-)relativistic velocity region. Besides, we find that the flux of ADM is insensitive to the masses of the mediator and dark matter when they are produced on-shell. This is because that the DM produced from $\eta$ meson decay has the kinetic energy $T_\chi=T_\eta \cdot \frac{m^2_\eta+m^2_S-m^2_\pi}{4m^2_\eta} \simeq \frac{T_\eta}{4}$. Since the kinetic energy of $\eta$ meson is strongly correlated with the power law of the primary CRs~\cite{Boschini:2017fxq}, it will produce a peak of $T_\eta$ at $\sim 2$ GeV, and then lead to a peak of $T_\chi$ at $\sim 0.5$ GeV.


\begin{figure}[ht]
  \centering
  \includegraphics[height=8cm,width=8.5cm]{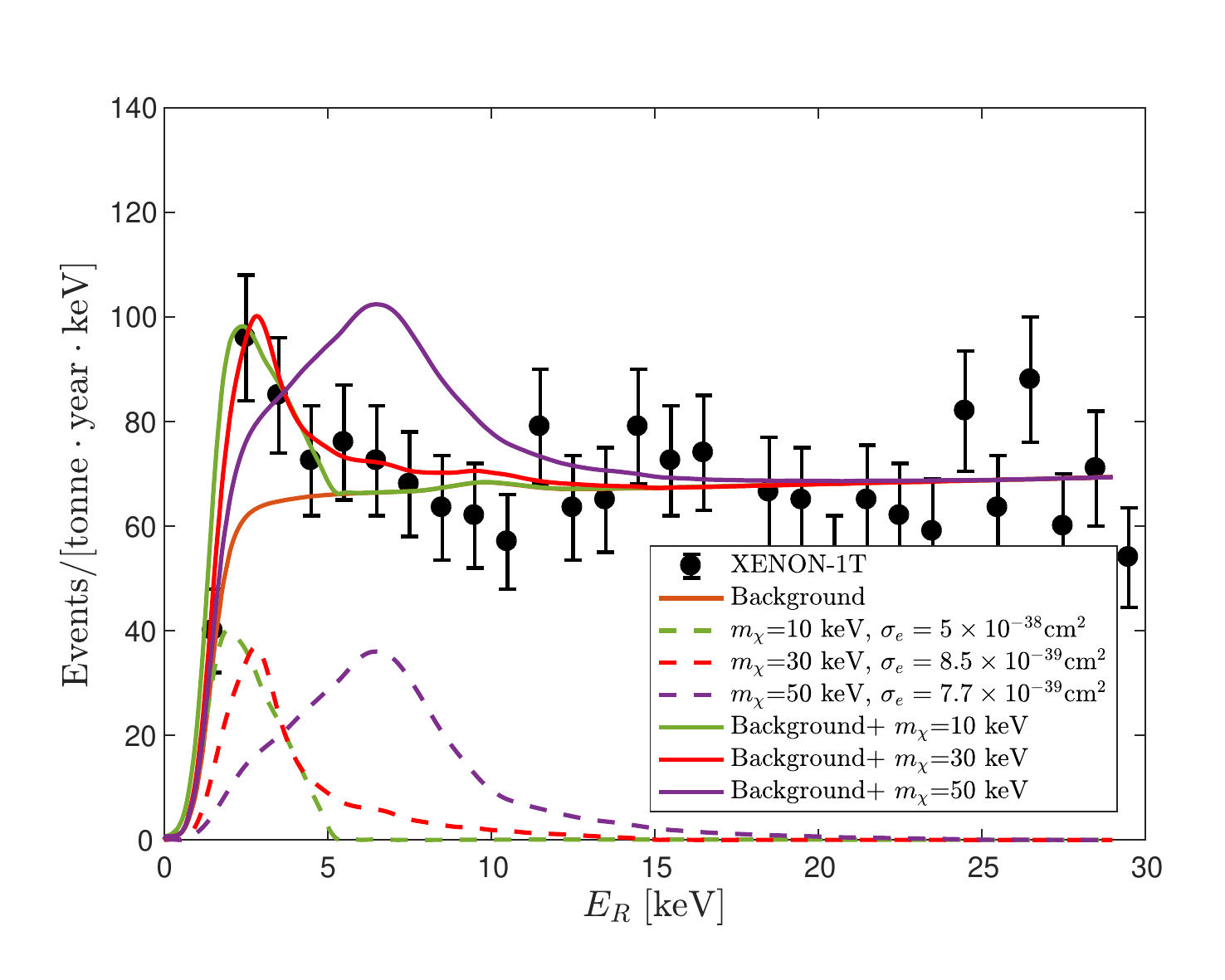}
  \caption{Same as Fig.~\ref{flux}, but for the ADM fit of Xenon1T electron recoil data. The signal and background (no tritium contribution) are plotted in yellow dashed line and purple solid line, respectively. The Xenon1T data points are taken from~\cite{Aprile:2020tmw} and are shown in black.}\label{spectrum}
\end{figure}

In Fig.~\ref{spectrum}, we present a fit of ADM to the Xenon1T electron recoil data by using Eq.~\ref{rate}. Thanks to the boosted effect of the CRs, our ADM can be energetic and produce the observed signal. From Eq.~\ref{cxer}, we note that the tendency of recoil spectrum is mainly affected by the mass and velocity of ADM (see also e.g. a model-independent study in Ref.~\cite{Kannike:2020agf}).
From the above discussions, we have known that the velocity of ADM is determined by the primary CRs, and thus the limits of integration of the momentum transfer $q_{\rm min}$ and $q_{\rm max}$ in Eq.~\ref{qtransfer} are only dependent on the mass of ADM for a given recoil energy $E_R$. The form factor $F(q)$ and ionization factor $f(q, E_R)$ can change the number of events but only slightly affect the position of the peak. By varying the values of DM mass $m_\chi$ and the DM-free electron scattering cross section $\sigma_e$, we find that the ADM with a mass of $10 \sim 30$ keV can fit the whole observed spectrum well. A heavier ADM will result in the deviation of predicted spectrum from the data.

\begin{figure}[ht]
\centering
\includegraphics[height=8cm,width=8.5cm]{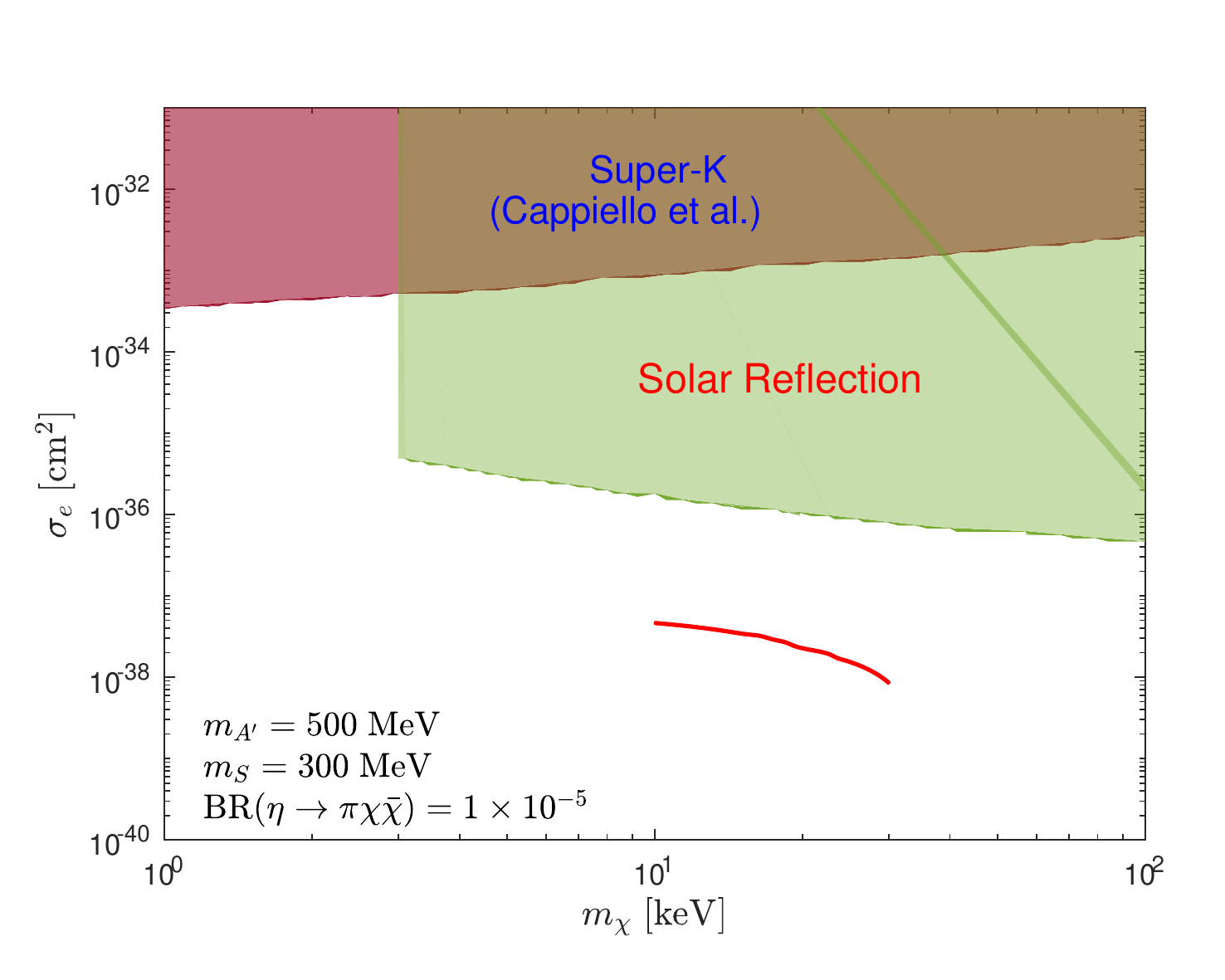}
\caption{The ADM-electron scattering cross section versus the mass of ADM (red curve). The exclusion limits from Super-Kamiokande neutrino experiment~\cite{Cappiello:2019qsw} and solar reflection~\cite{An:2017ojc} are also shown.}
\label{cx}
\end{figure}

In Fig.~\ref{cx}, we present the dependence of ADM-electron scattering cross section on the mass of ADM, in which each point on the red curve is required to fit the electron recoil spectrum in Xenon1T. Besides, we also show the exclusion limits from the Super-Kamiokande neutrino experiment and the solar reflection. It can be seen that our ADM-electron scattering cross sections vary from ${\cal O}(10^{-38}$) cm$^{2}$ to $\sim {\cal O}(10^{-39}$) cm$^{2}$, which are much smaller than those exclusion limits. Since our model for Xenon1T excess is highly predictive, the Xenon-nT may be able to test it in future. On the other hand, we should mention that our ADM-electron scattering cross section is also sensitive to the branching ratio of $\eta$ meson decay. Therefore, we can expect the future precision measurements of the light meson rare decay $\eta \to \pi + \slashed E_T$ in the low energy experiments would provide a complementary probe of our ADM.

Finally, we discuss the possible way to achieve the correct DM relic density in our scenario. As known, the thermal freeze-out DM in keV range is essentially ruled out, although some exceptions exist~\cite{Green:2017ybv,Berlin:2017ftj}. On the other hand, such a light DM can be produced from freeze-in~\cite{Hall:2009bx} as well. With the calculations in Ref.~\cite{Chang:2019xva}, we find that the correct relic density for a $10-30$ keV ADM requires $\sigma_e$ to be around $10^{-46}\mathrm{cm}^2$, which is below the value needed for Xenon1T anomaly. In other words, if we insist in explaining the Xenon1T excess, the corresponding relic density of DM will be larger than the experimental value. However, this paradox can be solved if the dark sector is diluted during the cosmological evolution~\cite{Evans:2019jcs}. The dilution may be produced by the decay of a heavy state, such as a long-lived moduli field or a messenger field. When the mediator is much heavier than the DM, the entropy injected into the SM will not feed back again. This mechanism has been proposed to solve the gravitino problem~\cite{Fujii:2002fv,Baltz:2001rq} and Hubble tension~\cite{Gu:2020ozv}, and then is extended to the vector portal model~\cite{Evans:2019jcs}. We leave the detailed study in our future work.

\section{Conclusions}
The very recent Xenon1T electron recoil excess in the keV range may be the evidence of the light dark matter. We proposed the atmospheric dark matter from the inelastic collision of cosmic rays with the atmosphere to interpret this excess. Due to the acceleration effect from high energy cosmic rays, we found that the light ADM can obtain enough kinetic energy and successfully fit the observed electron recoil spectrum via scattering with the electron. Besides, our ADM can also evade other direct detection constraints because of the momentum-dependent relativistic atomic form factor. On the other hand, since the ADM is produced from the meson decay in the cosmic-ray shower, the precision measurements of the light meson exotic decay $\eta \to \pi + \slashed E_T$ can test our ADM in future.

\section{acknowledgments}
We thank the helpful discussions with T.-T. Yu. This work is supported by the National Natural Science Foundation of China (NNSFC) under grant Nos. 117050934, 11775012, 11847208, 11875179, 11805161,  11675242, 11821505, 11851303,  by Jiangsu Specially Appointed Professor Program, by Peng-Huan-Wu Theoretical Physics Innovation Center (11847612),  by the CAS Center for Excellence in Particle Physics (CCEPP),  by the CAS Key Research Program of Frontier Sciences and by a Key R\&D Program of Ministry of Science and Technology under number 2017YFA0402204 and Natural Science Foundation of Shandong Province under the grants ZR2018QA007. BZ is also supported by the Basic Science Research Program through the National Research Foundation of Korea (NRF) funded by the Ministry of Education, Science and Technology (NRF-2019R1A2C2003738), and by the Korea Research Fellowship Program through the NRF funded by the Ministry of Science and ICT (2019H1D3A1A01070937).
\bibliography{refs}

 \end{document}